\documentclass[lettersize,journal]{IEEEtran}
\usepackage{amsmath,amsfonts}
\usepackage{amssymb}
\usepackage{algorithmic}
\usepackage{algorithm}
\usepackage{array}
\usepackage{textcomp}
\usepackage{stfloats}
\usepackage{url}
\usepackage{verbatim}
\usepackage{graphicx}
\usepackage{cite}
\usepackage{xcolor}
\usepackage[list=true]{subcaption}
\begin{document}

\title{RIS, Active RIS or RDARS: A Comparative Insight Through the Lens of Energy Efficiency}  
    \author{Aparna V C$^{1}$\thanks{1. Aparna V C and Sheetal Kalyani are with the Department of Electrical Engineering, Indian Institute of Technology, Madras, Chennai, India 600036. (Emails: \{ee22d026@smail, skalyani@ee\}.iitm.ac.in).}
    , Shashank Shekhar$^{2}$\thanks{2. Shashank Shekhar is with the Department of Electronics and Communication Engineering, PSG iTech, Coimbatore, India 641062. (Email: shashank@psgitech.ac.in)}, Sheetal Kalyani$^{1}$
    }    

\maketitle

\begin{abstract}
Multiplicative fading is a major limitation of reconfigurable intelligent surfaces (RIS), restricting their effective coverage in both existing sub-6GHz systems and future mmWave networks. Although active RIS architectures mitigate this issue, they require high power consumption and introduce practical challenges due to the need for integrated amplifiers. Recently, reconfigurable distributed antenna and reflecting surfaces (RDARS) have been proposed to alleviate multiplicative fading through connected modes. In this work, we compare RIS, active RIS, and RDARS in terms of coverage and energy efficiency (EE) in both sub-6GHz and mmWave bands, and we investigate the impact of placement and the number of elements of reconfigurable surface (RS) on EE and coverage. The simulation results show that RDARS offers a highly energy-efficient alternative of enhancing coverage in sub-6GHz systems, while active RIS is significantly more energy-efficient in mmWave systems. Additionally, for a lower number of RS elements and for near UEs, RIS remains considerably more energy-efficient than both active RIS and RDARS.
\end{abstract}

\begin{IEEEkeywords}
reconfigurable intelligent surfaces, reconfigurable distributed antenna and reflecting surfaces, active RIS, energy efficiency, sub-6GHz, mmWave
\end{IEEEkeywords}

\section{Introduction}
\IEEEPARstart{R}{econfigurable} Intelligent Surface (RIS) is a technology that enable the wireless propagation environment to be programmable by adjusting the phase shifts of impinging signals\cite{9140329,8801961}. Although RIS is a cost-effective and energy-efficient solution, a fundamental issue is the multiplicative fading, since the signal experiences path loss on both the transmitter–RIS and RIS–receiver links, the overall attenuation increases. This creates major challenges to the application of RIS in the existing 5G systems operating in the sub-6GHz band and in the future mmWave based 6G systems also.

Active RIS has been proposed to overcome the multiplicative fading effect \cite{9377648,zhang2022active,9734027}. However, it requires significantly higher power consumption and introduces practical implementation challenges, as each active RIS element must be integrated with an amplifier to boost and reflect the signal. Recently, reconfigurable distributed antenna and reflecting surfaces (RDARS) has been proposed as a solution to address the multiplicative fading problem by providing connected modes between the RDARS and the base station (BS) \cite{ma2024reconfigurable}.

Although prior works have extensively studied RIS and active RIS \cite{9140329,8801961,charishma2021outage,9377648}, and some works have compared RIS with active RIS \cite{zhang2022active,9734027}, a unified comparison that encompasses all three architectures passive RIS, active RIS, and RDARS is missing in the literature. This research gap motivated us to systematically evaluate their performance, delineate their respective strengths, and identify the operating conditions under which each architecture is most effective. Given that contemporary 5G and Wi-Fi systems predominantly operate in the sub-6GHz band, we first provide a comprehensive comparison of the three architectures in this frequency range, with emphasis on coverage enhancement and energy efficiency (EE). We further extend our investigation to the mmWave regime. Owing to the severe path loss and attenuation at these frequencies, conventional passive RIS often becomes ineffective. Accordingly, we examine whether active RIS or RDARS can mitigate these challenges and assess the extent to which optimized placement can improve coverage and EE in mmWave systems.

In summary, this work presents, to the best of our knowledge, the first unified comparative analysis of passive RIS, active RIS, and RDARS across both sub-6GHz and mmWave bands based on multiple metrics such as received signal to noise ratio (SNR), EE with respect to various system parameters such as number of elements and placement of reconfigurable surface (RS). Our objective is to elucidate their relative advantages, limitations, and trade-offs, highlighting that no single architecture is universally optimal.

\section{System Model}
Consider a communication system with a single-antenna user equipment (UE), a RS with $N$ elements, where RS can be RIS or active RIS or RDARS, and a single-antenna BS. The channel coefficients between UE-BS, UE-RS, and RS-BS are represented by $h_{UB} \in \mathbb{C},\ \mathbf{h}_{UR},\mathbf{h}_{RB} \in \mathbb{C}^{N \times 1}$ where ${h}_{RB_{i}}$ and ${h}_{UR_{i}}$ are the $i$-th element of $\mathbf{h}_{RB}$ and $\mathbf{h}_{UR}$, respectively. We considered the uplink scenario where $x \in \mathbb{C}$ is the transmitted signal from UE to BS with total power $\mathbb{E}\left[|x|^2\right] \leq P_{t}$. The working of RDARS and active RIS-assisted communication system is presented in the subsequent subsection along with their corresponding SNR expressions at the receiver. However, conventional RIS system can be viewed as a special case of RDARS system. 

\subsection{RDARS}
The system model for RDARS-assisted system is adopted from \cite{ma2024reconfigurable}. The total number of elements in the RDARS is $N$, and each element can operate either in connected mode or in reflection mode. 
In connected mode, an element functions as a remote antenna linked to the BS via dedicated wires or fibers. In reflection mode, it behaves as a passive reflecting unit similar to a conventional RIS element. The number of elements in connected mode denoted by $a$, are much fewer than the number of elements, \textit{i.e.}, $a \ll N$.
 
The RDARS configuration is characterized by a mode-indicating matrix $\mathbf{A} \in \mathbb{C}^{a \times N}$, where $\mathbf{A}^H\mathbf{A}(i,i) \in \{0,1\}$. Specifically, $\mathbf{A}^H\mathbf{A}(i,i)=1$ indicates that the $i$th element operates in connected mode, while $\mathbf{A}^H\mathbf{A}(i,i)=0$ denotes reflection mode, in which the element only applies a phase shift to the incoming signal. The reflection-coefficient matrix of RDARS is given by
\[
    \mathbf{B} \triangleq (\mathbf{I}-\mathbf{A}^H\mathbf{A})\boldsymbol{\Theta}
    \in \mathbb{C}^{N \times N}.
\]
where $\boldsymbol{\Theta}=\mathrm{diag}(\boldsymbol{\theta}) \in \mathbb{C}^{N \times N}$ is the phase shift matrix, with ${\boldsymbol\Theta}(i,i)={\theta_{i}}=e^{j\arg({\theta_{i}})}$, $\theta_{i}$ is the $i$-th element of vector $\boldsymbol{\theta}$, and $\arg({\theta_{i}})$ is the phase shift induced by the $i$-th element.

Due to the presence of connected modes, the BS will receive signal not only through the antenna but also through the connected elements of RDARS. The combined received signal at the BS is as follows
\begin{equation}
    \begin{aligned}
        \begin{bmatrix} y \\ \mathbf{u} \end{bmatrix} = 
        \begin{bmatrix}
            h_{UB} + \mathbf{h}_{RB}^T\mathbf{B}\mathbf{h}_{UR} \\ \mathbf{Ah}_{UR}
        \end{bmatrix} x + 
        \begin{bmatrix}
            n_{BS} \\ \mathbf{n}_R
        \end{bmatrix}
    \end{aligned}
\end{equation}
where, $y \in \mathbb{C}$ is the received signal at the antenna and $\mathbf{u} \in \mathbb{C}^{a \times 1}$ is the received signal through the connected modes, $n_{BS} \sim \mathcal{CN}\left(0, \sigma_{B S}^2\right)$ and $n_{R} \sim \mathcal{CN}\left(0, \sigma_{R}^2\mathbf{I}\right)$ are the thermal noise component of the aforementioned signal component, respectively. The BS and RDARS noises are assumed to have equal power \textit{i.e.}, $\sigma_R^2 = \sigma_{BS}^2 = \sigma^2$ \cite{ma2024reconfigurable}. The received SNR ($\gamma_{RDARS}$) after maximum ratio combining (MRC) is given by
    \begin{equation}
    \gamma_{RDARS} = \overline{\gamma}\Big(\left|h_{UB}+\mathbf{h}_{R B}^T \mathbf{B h}_{U R}\right|^2+\mathbf{h}_{U R}^H \mathbf{A}^H \mathbf{A h}_{U R}\Big)
    \end{equation}
where $\overline{\gamma}=P_{t}/\sigma^2$ is the transmit SNR. For elements operating in reflection mode in a RDARS, optimal phase shifts are given as \cite{ma2024reconfigurable}
    \begin{equation}
    \arg(\theta_{i}) = \arg(h_{UB})-\arg({h}_{RB_{i}}{h}_{UR_{i}}), \forall \mathbf{A}^{H}\mathbf{A}(i,i)=0 
    \end{equation}
With optimal phase shifts, the received SNR is simplified as 
    \begin{multline}
    \begin{aligned}
    \gamma_{RDARS} = \overline{\gamma} \Bigg[\Big(\left|h_{U B}\right|+ \sum_{i=1}^{N} \left(1-a_{i}\right)\left| {h}_{RB_{i}}\right| \left|{h}_{UR_{i}}\right|\Big)^2 \\
    + \mathbf{h}_{U R}^H \mathbf{A}^H \mathbf{A h}_{U R}\Bigg] 
    \end{aligned}
    \end{multline}
    where $a_i = \mathbf{A}^{H}\mathbf{A}(i,i) \in \{0,1\}$.
    
\subsection{Active RIS}
In an active RIS with $N$ elements, each element amplifies and applies a phase shift to the impinging signal. However, this amplification also increases the noise forwarded to the receiver. We followed the system model presented in \cite{9377648} which is widely adopted in the literature. Accordingly, the received signal at the BS in an active RIS-assisted system is
    \begin{multline}
    \begin{aligned}
    y &=h_{UB}x+\mathbf{h}_{RB} \boldsymbol{\Theta}\left(\mathbf{h}_{UR}x+\mathbf{n}_{2}\right)+n_{1} \\
    &=\left(h_{UB}+\mathbf{h}_{RB} \boldsymbol{\Theta} \mathbf{h}_{UR}\right)x +\mathbf{h}_{RB} \boldsymbol{\Theta} \mathbf{n}_{2}+n_{1}
    \end{aligned}
    \end{multline}          
where $n_{1} \sim \mathcal{C N}\left(0, \sigma_{1}^{2} \right)$ and $\mathbf{n}_{2} \sim \mathcal{C N}\left(\mathbf{0}, \sigma_{2}^{2} \mathbf{I}_{N}\right)$ represent the thermal noise at the receiver and the active RIS, respectively. The second term in the above equation is usually neglected as it is relatively small compared to the noise power at the receiver. However active RIS amplify the incident signal, so the noise introduced at the active RIS can no longer be ignored \cite{9377648}. Hence, the received SNR $\gamma_{ARIS}$ after MRC is given by ,
    \begin{equation}
    \gamma_{ARIS}=\frac{P_{t}\left|\mathbf{h}_{UB}+\mathbf{h}_{RB}^T \boldsymbol{\Theta} \mathbf{h}_{UR}\right|^{2}}{\sigma_{2}^{2}\left\|\mathbf{h}_{RB}^T \boldsymbol{\Theta}\right\|^{2}+\sigma_{1}^{2}}    
    \end{equation}
    
 The optimal reflecting coefficients for each element in the active RIS that maximizes $\gamma_{ARIS}$ are \cite{9377648}
    \begin{equation}
    \begin{aligned}
     \arg(\theta_{i}) = \frac{\sigma_1^2\left|h_{UR_{i}}\right|}{\sigma_2^2\left|h_{UB}\right|\left|h_{RB_{i}}\right|}\big(\arg(h_{UB})-\arg(h_{UR_{i}}h_{RB_{i}})\big)   
    \end{aligned}     
    \end{equation}    

With optimal phase shifts and assuming that the power of noises at the receiver ($\sigma_{1}^2$) and active RIS ($\sigma_{2}^2$) are the same \cite{9377648}, the received SNR is simplified as 
    \begin{equation}
         \gamma_{ARIS}= \overline{\gamma} \left(\left|h_{UB}\right|^2 +\left\|\mathbf{h}_{UR}\right\|^2\right)
    \end{equation}   
    
\section{Power Consumption Model and Energy Efficiency}
\subsection{Power Consumption Model}
The total power consumed in a RDARS system is given by 
\begin{equation}
\label{P_RDARS}
    P_{RDARS} = (N-a)P_C+aP_{RF} 
\end{equation}
where $N$ is the number of RDARS elements, $a$ is the number of connected modes, $P_C$ is the power of the switching and control circuit and $P_{RF}$ is the power consumed by the RF chain associated with the connected modes. The power consumed by the RF chain at mmWave and sub-6GHz is adopted from \cite{abbas2017millimeter,hashemi2017energy}. The total power consumed by RIS is $NP_{C}$ as there are no connected modes in the conventional RIS.

Similarly, the total power consumed by the active RIS is defined as 
\begin{equation}
\label{P_ARIS}
    P_{ARIS} = N(P_C+P_{DC})+\zeta p_{out}    
\end{equation}
where $P_{DC}$ is the DC biasing power required for each element in active RIS for amplification. In the above equation $\zeta p_{out}$ is very small and hence ignored for the calculation of the total power consumed in an active RIS \cite{9377648}. The values of $P_{C}$ and $P_{DC}$ are adopted from \cite{9377648,wang2024reconfigurable}.

Since $P_{c}$ is very small, the power consumption of active RIS and RDARS is dominated by $P_{DC}$ and $P_{RF}$, respectively. As the number of elements ($N$) increases, the power consumption of the active RIS increases, whereas the power consumption of RDARS does not change significantly as $P_{RF}$ is independent of $N$.

\subsection{Energy Efficiency}
Energy efficiency (EE) is a metric \cite{ngo2017total} which is defined as the ratio of capacity to total power consumed and is given by
\begin{equation}
    EE = B\log_2(1+\gamma)/P
\end{equation}
where $\gamma$ is the received SNR, $B$ is the bandwidth and $P$ is the total power consumed by the respective device.

\section{Simulation Results}
In this section, we compare the performance of RIS, RDARS, and active RIS systems at both sub-6GHz and mmWave frequencies. The system layout follows \cite{ma2024reconfigurable}, where the BS, RS, and UE are located at (0m, 0m, 10m), (20m, 20m, 10m), and (200m, 0m,1.5m), respectively. The number of reflecting elements ($N$) is set to 1024, and the number of elements in the connected mode ($a$) is varied from 1 to 4 \cite{ma2024reconfigurable}. The path-loss models are adopted from the latest 3GPP standards \cite{3gpp38901}. The channel model and simulation parameters for sub-6GHz are taken from \cite{ma2024reconfigurable}, while those for mmWave are taken from \cite{raj2022deep,8207426}. All simulation results are obtained by averaging over 30,000 channel realizations. 

The parameters used in the simulation are summarized in Table \ref{Dtab1}. 
\begin{table}[!h]
    \begin{center}
    \caption{Simulation Parameters}
    \label{Dtab1}
    \begin{tabular}{|c|c|c|} 
        \hline
        \textbf{Parameters} & \textbf{sub-6GHz}\cite{ma2024reconfigurable} & \textbf{mmWave}\cite{8207426} \\
        \hline
        Frequency (GHz) & 3.7 & 28 \\
        \hline
        Transmit Power (dBm) & 10 & 20 \\
        \hline
        Bandwidth (MHz) & 20 & 500 \\ 
        \hline
        Noise Power (dBm) & -100 & -87 \\ 
        \hline
    \end{tabular}
    \end{center}
\end{table}

\begin{figure}[!h]
     \centering
     \begin{subfigure}[b]{0.475\textwidth}
         \centering
          \includegraphics[width=\textwidth]{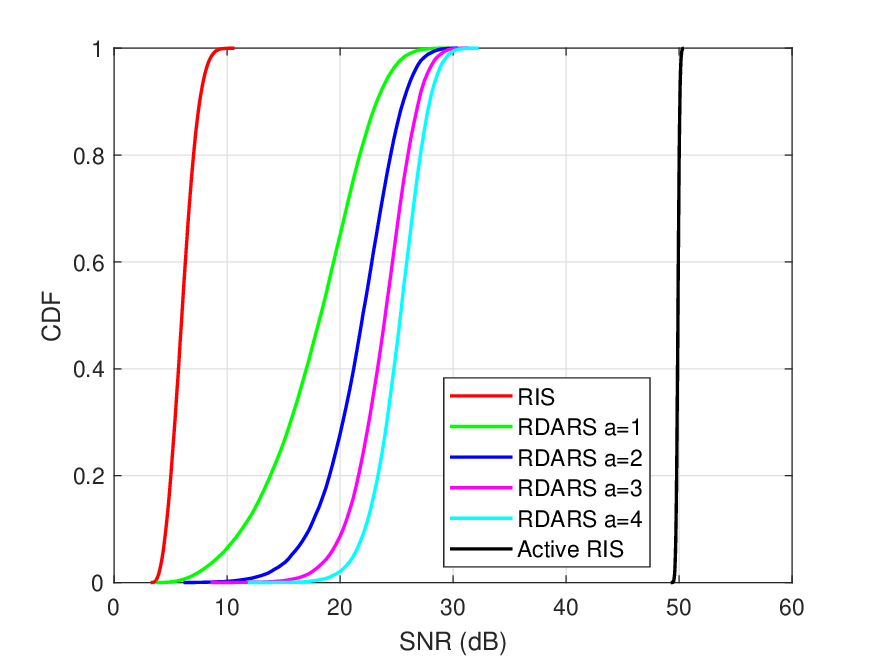}     
         \caption{sub-6GHz}
         \label{cdf_sub}
     \end{subfigure}
     \hfill
     \begin{subfigure}[b]{0.475\textwidth}
         \centering
         \includegraphics[width=\textwidth]{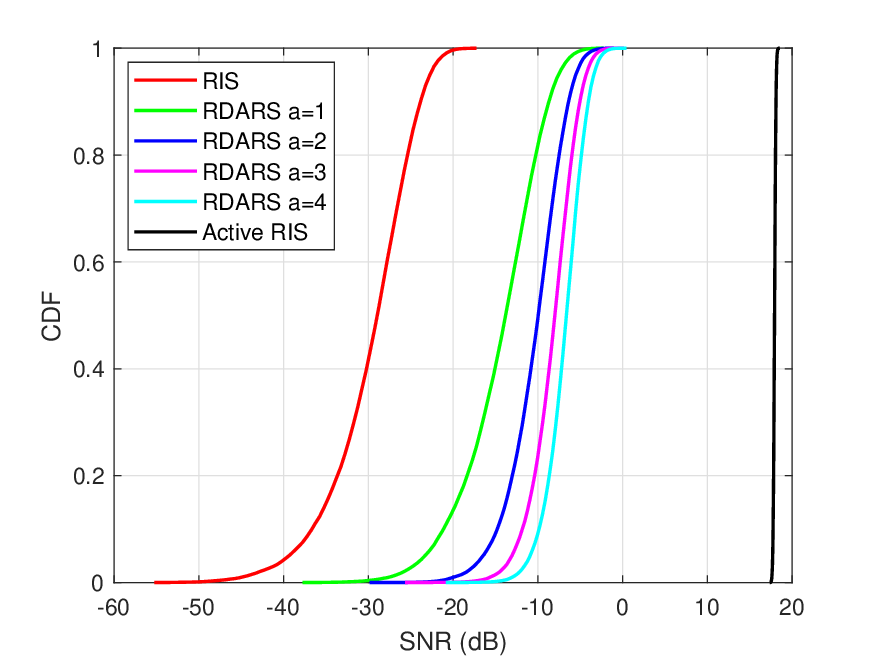}       
         \caption{mmWave}
         \label{cdf_mm}
     \end{subfigure}
     \hfill
     \caption{CDF of the received SNR}
     \label{cdf}
\end{figure}

In Fig. \ref{cdf}, we plot the Cumulative Distribution Function (CDF) of the received SNR for RIS, RDARS, and active RIS systems operating at both sub-6GHz and mmWave frequencies. The RDARS system with single connected mode exhibits a significant improvement in SNR compared to the conventional RIS system at both sub-6GHz and mmWave. Furthermore, as the number of connected modes ($a$) increases, the SNR also improves. By leveraging connected modes, RDARS effectively mitigates the multiplicative fading that constitutes the fundamental bottleneck of RIS. At mmWave frequencies, the performance of RIS deteriorates further due to the combined effects of higher path loss and multiplicative fading. In contrast, the active RIS clearly outperforms both RIS and RDARS at sub-6GHz and mmWave bands, but this superior performance comes at the expense of higher power consumption and circuit complexity. So in the subsequent simulation results we will try to investigate the performance of three devices using EE metric. This metric emphasizes the importance of evaluating not just performance but also the energy cost at which it is achieved.

\begin{figure}[!h]
     \centering
     \begin{subfigure}[b]{0.475\textwidth}
         \centering
          \includegraphics[width=\textwidth]{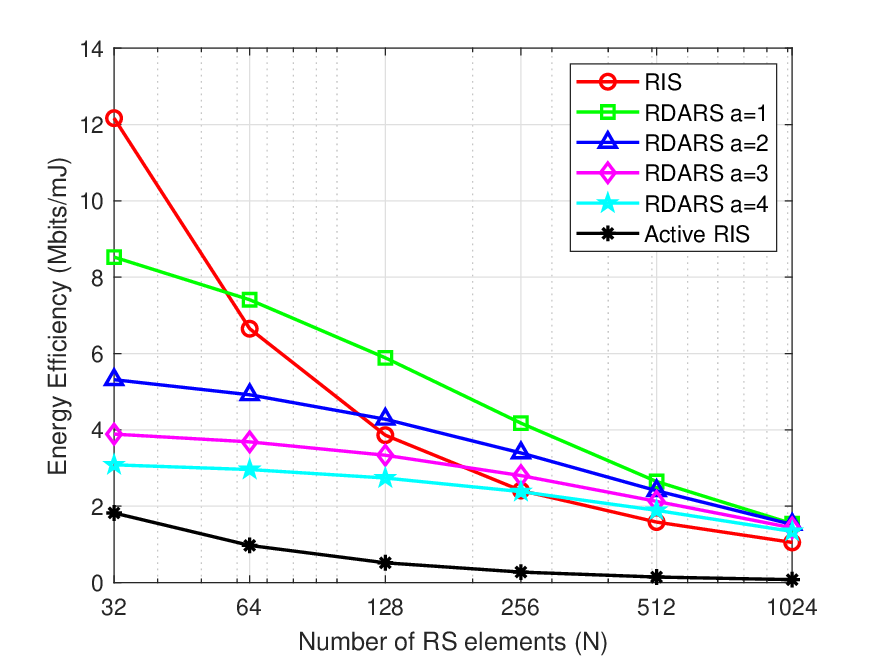}      
         \caption{sub-6GHz}
         \label{ee_N_sub}
     \end{subfigure}
     \hfill
     \begin{subfigure}[b]{0.475\textwidth}
         \centering
         \includegraphics[width=\textwidth]{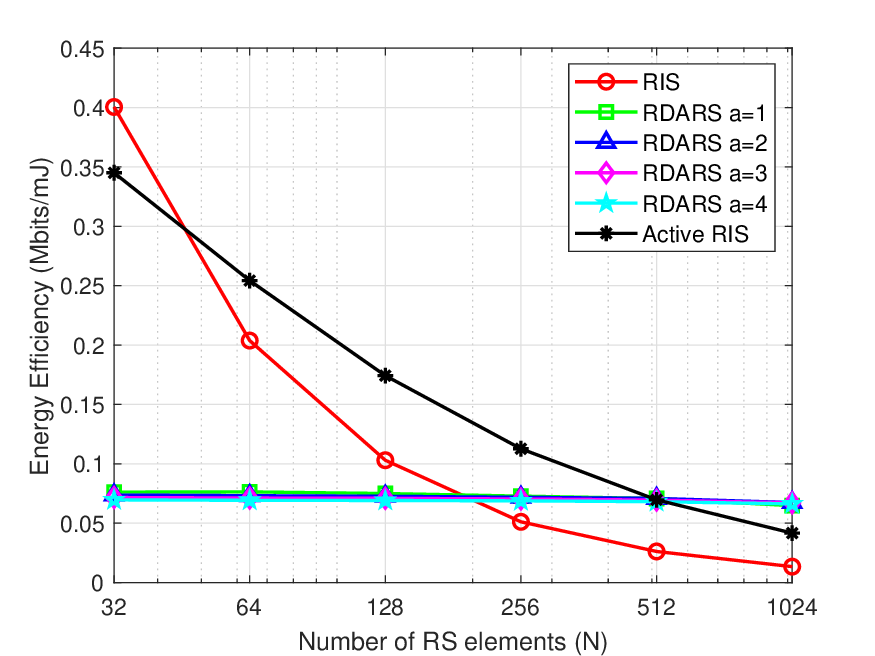}        
         \caption{mmWave}
         \label{ee_N_mm}
     \end{subfigure}
     \hfill     
     \caption{EE vs. Number of RS elements ($N$)}
     \label{ee_N}
\end{figure}

Since, power consumption and received SNR both increases with the number of reflecting elements, it is important to analyze how the EE varies with $N$ to understand the cost of performance improvement. From Fig. \ref{ee_N}, we observe that the EE of both RIS and active RIS decreases with increasing $N$ in both sub-6GHz and mmWave systems. For RDARS, EE decreases with $N$ in the sub-6GHz band, but it remains nearly constant in the mmWave regime. This behavior arises because the RDARS power model depends on both $N$ and the number of connected-mode elements ($a$). The RF-chain power associated with the connected-mode elements dominates over that of the reflected-mode elements ($N-a$), particularly at mmWave frequencies. Consequently, the EE of RDARS is insensitive to $N$ in mmWave.

Interestingly, in the sub-6GHz band, RIS consistently achieves higher EE than active RIS for all $N$, whereas in the mmWave band, active RIS outperforms RIS except when $N \leq 40$. For small $N$, RIS is the most energy-efficient scheme in both sub-6GHz and mmWave bands. For large $N$, RDARS achieves higher EE than both RIS and active RIS in both bands. At moderate values of $N$, RDARS outperforms RIS and active RIS in sub-6GHz, while active RIS attains the highest EE in the mmWave regime. Furthermore, as shown in the subsequent plot, proper placement of the RS enables RDARS to surpass both RIS and active RIS under certain conditions.

\begin{figure}[!h]
     \centering
     \begin{subfigure}[b]{0.475\textwidth}
         \centering        
         \includegraphics[width=\textwidth]{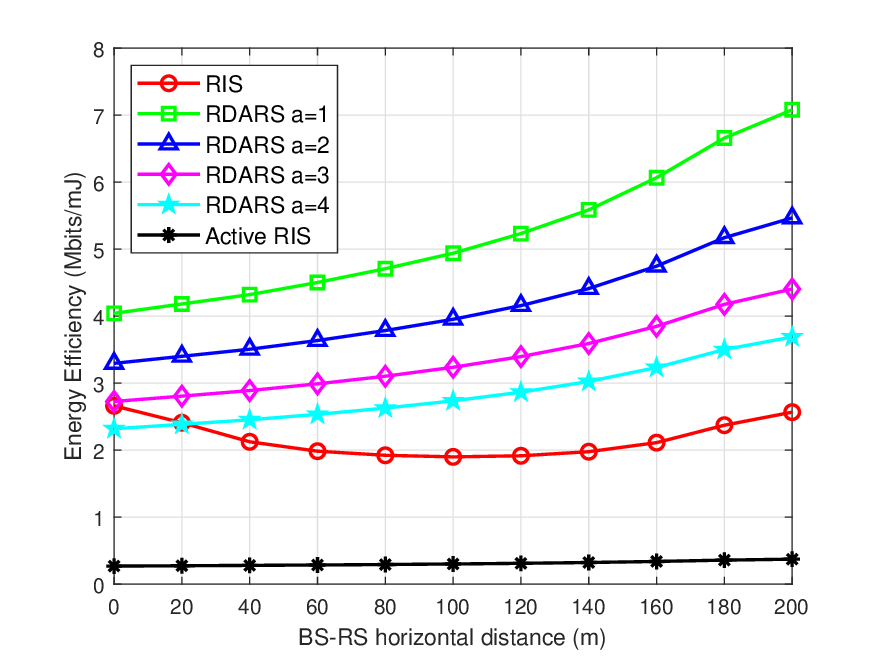}
         \caption{sub-6GHz}
         \label{d_sub}
     \end{subfigure}
     \hfill
     \begin{subfigure}[b]{0.475\textwidth}
         \centering
         \includegraphics[width=\textwidth]{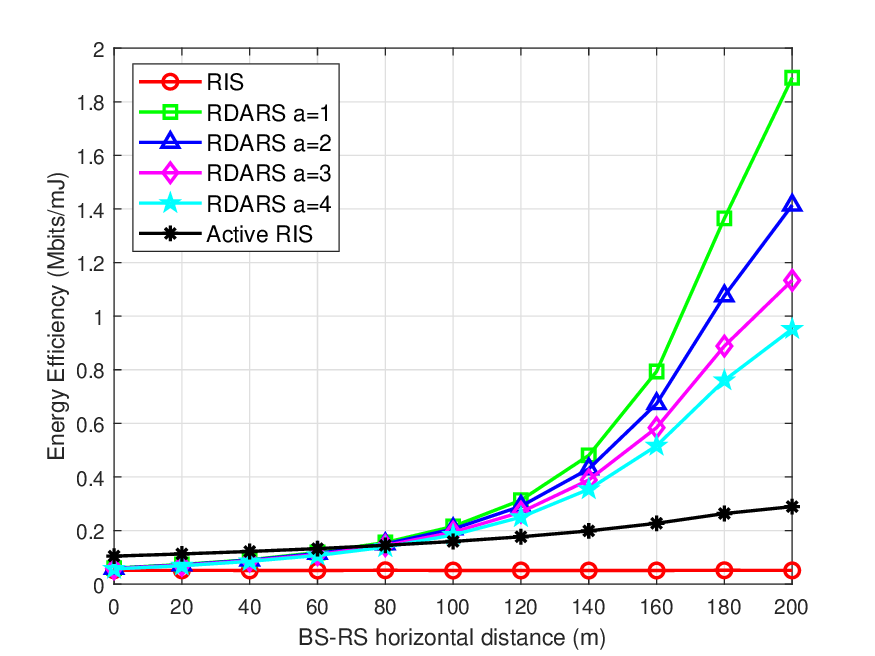}
         \caption{mmWave}
         \label{d_mm}
     \end{subfigure}
     \hfill
     \caption{EE vs. BS-RS horizontal distance, $N=256$}
     \label{d}
\end{figure}

It is observed from Fig. \ref{ee_N} that EE of RIS is monotonic with $N$ and it has cross-over with EE of RDARS at $N=256$, so it is interesting to investigate the impact of BS–RS horizontal distance on EE for $N=256$. Hence, in Fig. \ref{d}, the EE is plotted against the BS–RS horizontal distance ($x_{RS}$) for $N=256$. In the sub-6GHz band, the EE of RDARS with $a$ varied from 1 to 4 is higher than that of the active RIS for all values of $x_{RS}$. As $a$ increases, the BS requires more RF chains, which raises its power consumption and consequently reduces the EE of RDARS. Nevertheless, RDARS with $a$ between 1 and 3 still achieves higher EE than both the active RIS and the passive RIS across all distances. As $x_{RS}$ increases, the EE of RDARS improves in both sub-6GHz and mmWave bands because placing RDARS closer to the UE reduces the effective RS–UE distance and hence the path loss, while the connected mode helps mitigate fading between the BS and RDARS. For example: in the mmWave regime, for $x_{RS}\leq60 \hspace{1mm}m$, RDARS has poor EE compared to active RIS and for $x_{RS}\geq100$, EE of RDARS is higher than that of both the active RIS and the passive RIS. However, placing RDARS very close to the UE introduces practical challenges because a physical connection between the BS and RDARS is required.   

Summarizing, the EE of RDARS either decreases (in sub-6GHz) or remains almost constant (in mmWave) with $N$ and it monotonically increases with $x_{RS}$. So, for any fixed value of $N$, one can increase $x_{RS}$ to get a better value of EE if it is possible to maintain the physical connection between BS and RDARS at that $x_{RS}$.

\begin{figure}[!h]
     \centering
     \begin{subfigure}[b]{0.475\textwidth}
         \centering
         \includegraphics[width=\textwidth]{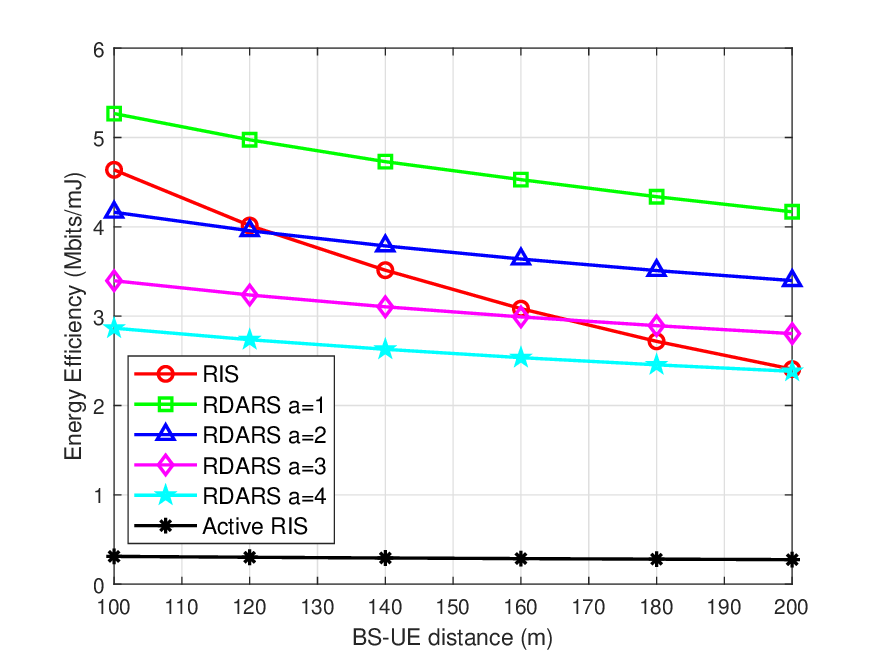}
         \caption{sub-6GHz}
         \label{ue_sub}
     \end{subfigure}
     \hfill
     \begin{subfigure}[b]{0.475\textwidth}
         \centering
         \includegraphics[width=\textwidth]{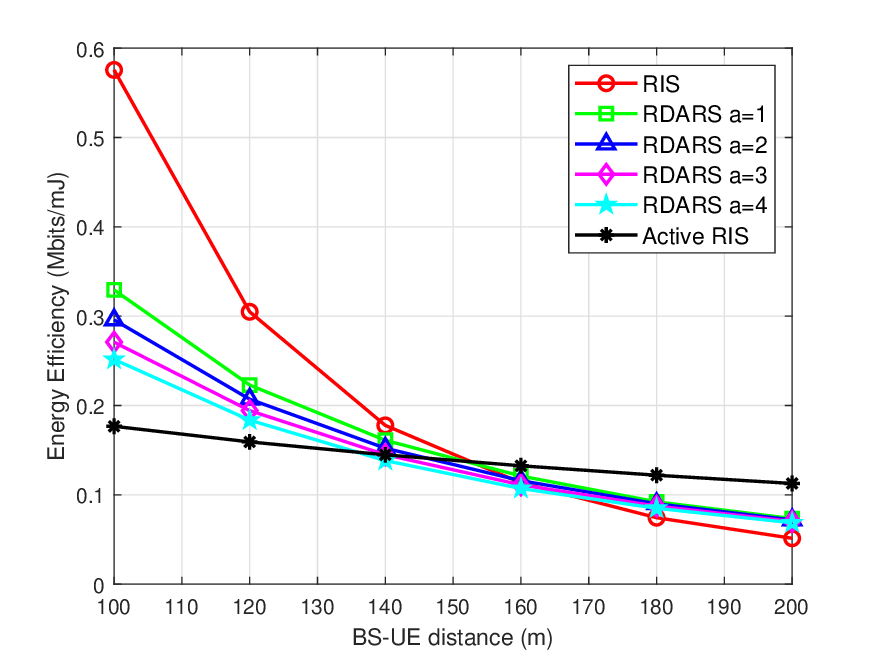}
         \caption{mmWave}
         \label{ue_mm}
     \end{subfigure}
     \hfill
     \caption{EE vs. BS-UE distance, $N=256$}
     \label{ue}
\end{figure} 

Finally, we want to study the impact of BS–UE distance on the EE. In Fig. \ref{ue}, the EE of RIS, RDARS, and active RIS is plotted as a function of the BS–UE distance in both the sub-6GHz and mmWave bands for $N = 256$. In the sub-6GHz band, the EE of RDARS and RIS is significantly higher than that of active RIS. For nearby UEs, RIS is more energy-efficient than RDARS when $a\geq2$; however, for distant UEs, the EE of RDARS outperforms that of RIS. In the mmWave regime, RIS is more energy-efficient than both RDARS and active RIS for near UEs, but for distant UEs, the EE of active RIS surpasses that of RIS and RDARS. Although RDARS can ensure improved coverage with good EE in the sub-6GHz band, in the mmWave band active RIS becomes more energy-efficient and provides more robust coverage. However, as shown in Fig. \ref{d}, by suitably positioning the RDARS ($x_{RS}\geq100$), the EE of RDARS can be improved while still ensuring high coverage.

\section{Conclusion}
In this work, we compared the performance of RIS, RDARS, and active RIS at sub-6GHz and mmWave frequencies in terms of coverage and EE, and examined the effects of RS placement and the number of elements. The simulation results show that for small $N$ and nearby  users, RIS achieves the highest EE in the sub-6GHz band. For larger $N$, and when the RS is placed closer to the UE, RDARS outperforms both RIS and active RIS in EE and coverage, particularly in the mmWave band. Considering the practical limitations of positioning RDARS near the UE, active RIS becomes a favorable option in mmWave systems to achieve better coverage and higher EE. In contrast, in sub-6GHz systems, placing the RS close to the BS enables RDARS to achieve significantly higher EE than both RIS and active RIS while ensuring robust coverage.

\bibliographystyle{IEEEtran}
\bibliography{References}

@ARTICLE{9140329,
  author={Di Renzo, Marco and Zappone, Alessio and Debbah, Merouane and Alouini, Mohamed-Slim and Yuen, Chau and de Rosny, Julien and Tretyakov, Sergei},
  journal={IEEE Journal on Selected Areas in Communications}, 
  title={Smart Radio Environments Empowered by Reconfigurable Intelligent Surfaces: How It Works, State of Research, and The Road Ahead}, 
  year={2020},
  volume={38},
  number={11},
  pages={2450-2525},
  doi={10.1109/JSAC.2020.3007211}}

@INPROCEEDINGS{8801961,
  author={Basar, Ertugrul},
  booktitle={2019 European Conference on Networks and Communications (EuCNC)}, 
  title={Transmission Through Large Intelligent Surfaces: A New Frontier in Wireless Communications}, 
  year={2019},
  volume={},
  number={},
  pages={112-117},
  doi={10.1109/EuCNC.2019.8801961}}

@ARTICLE{9734027,
  author={Zhi, Kangda and Pan, Cunhua and Ren, Hong and Chai, Kok Keong and Elkashlan, Maged},
  journal={IEEE Communications Letters}, 
  title={Active {RIS} Versus Passive {RIS}: Which is Superior With the Same Power Budget?}, 
  year={2022},
  volume={26},
  number={5},
  pages={1150-1154},
  doi={10.1109/LCOMM.2022.3159525}}

@article{zhang2022active,
  title={Active {RIS} vs. passive {RIS}: Which will prevail in 6{G}?},
  author={Zhang, Zijian and Dai, Linglong and Chen, Xibi and Liu, Changhao and Yang, Fan and Schober, Robert and Poor, H Vincent},
  journal={IEEE Transactions on Communications},
  volume={71},
  number={3},
  pages={1707--1725},
  year={2022},
  publisher={IEEE}
}

@ARTICLE{9377648,
  author={Long, Ruizhe and Liang, Ying-Chang and Pei, Yiyang and Larsson, Erik G.},
  journal={IEEE Transactions on Wireless Communications}, 
  title={Active Reconfigurable Intelligent Surface-Aided Wireless Communications}, 
  year={2021},
  volume={20},
  number={8},
  pages={4962-4975},
  doi={10.1109/TWC.2021.3064024}}

@article{ma2024reconfigurable,
  title={Reconfigurable distributed antennas and reflecting surface: A new architecture for wireless communications},
  author={Ma, Chengzhi and Yang, Xi and Wang, Jintao and Yang, Guanghua and Zhang, Wei and Ma, Shaodan},
  journal={IEEE Transactions on Communications},
  volume={72},
  number={10},
  pages={6583--6598},
  year={2024},
  publisher={IEEE}
}

@article{charishma2021outage,
  title={Outage probability expressions for an {IRS}-assisted system with and without source-destination link for the case of quantized phase shifts in $\kappa$--$\mu$ fading},
  author={Charishma, Mavilla and Subhash, Athira and Shekhar, Shashank and Kalyani, Sheetal},
  journal={IEEE Transactions on Communications},
  volume={70},
  number={1},
  pages={101--117},
  year={2021},
  publisher={IEEE}
}

@article{abbas2017millimeter,
  title={Millimeter wave receiver efficiency: A comprehensive comparison of beamforming schemes with low resolution {ADC}s},
  author={Abbas, Waqas Bin and Gomez-Cuba, Felipe and Zorzi, Michele},
  journal={IEEE Transactions on Wireless Communications},
  volume={16},
  number={12},
  pages={8131--8146},
  year={2017},
  publisher={IEEE}
}

@article{hashemi2017energy,
  title={Energy-Efficient Power and Bandwidth Allocation in an Integrated {S}ub-6 {GH}z-Millimeter Wave System},
  author={Hashemi, Morteza and Koksal, C Emre and Shroff, Ness B},
  journal={arXiv preprint arXiv:1710.00980},
  year={2017}
}

@article{wang2024reconfigurable,
  title={Reconfigurable intelligent surface: Power consumption modeling and practical measurement validation},
  author={Wang, Jinghe and Tang, Wankai and Liang, Jing Cheng and Zhang, Lei and Dai, Jun Yan and Li, Xiao and Jin, Shi and Cheng, Qiang and Cui, Tie Jun},
  journal={IEEE Transactions on Communications},
  volume={72},
  number={9},
  pages={5720--5734},
  year={2024},
  publisher={IEEE}
}

@article{ngo2017total,
  title={On the total energy efficiency of cell-free massive {MIMO}},
  author={Ngo, Hien Quoc and Tran, Le-Nam and Duong, Trung Q and Matthaiou, Michail and Larsson, Erik G},
  journal={IEEE Transactions on Green Communications and Networking},
  volume={2},
  number={1},
  pages={25--39},
  year={2017},
  publisher={IEEE}
}

@article{raj2022deep,
  title={Deep Reinforcement Learning based blind mm{W}ave {MIMO} beam alignment},
  author={Raj, Vishnu and Nayak, Nancy and Kalyani, Sheetal},
  journal={IEEE Transactions on Wireless Communications},
  volume={21},
  number={10},
  pages={8772--8785},
  year={2022},
  publisher={IEEE}
}

@techreport{3gpp38901,
  title        = {{Study on Channel Model for Frequencies from 0.5 to 100 GHz}},
  institution  = {{3rd Generation Partnership Project (3GPP)}},
  type         = {{Technical Report}},
  number       = {{3GPP TR 38.901 V16.1.0}},
  year         = {2020},
  month        = {November},
  note         = {{Release 16}},
  url          = {https://www.3gpp.org/ftp/Specs/archive/38_series/38.901/38901-f10.zip}
}

@ARTICLE{8207426,
  author={Hemadeh, Ibrahim A. and Satyanarayana, Katla and El-Hajjar, Mohammed and Hanzo, Lajos},
  journal={IEEE Communications Surveys \& Tutorials}, 
  title={Millimeter-Wave Communications: Physical Channel Models, Design Considerations, Antenna Constructions, and Link-Budget}, 
  year={2018},
  volume={20},
  number={2},
  pages={870-913},
  doi={10.1109/COMST.2017.2783541}}

\end{document}